# Perfect Absorption in a Metasurface-Programmable Complex Scattering Enclosure

Mohammadreza F. Imani[1], David R. Smith[1], Philipp del Hougne[2*]

[1] Department of Electrical and Computer Engineering, Duke University, Durham, North Carolina 27708, USA

[2] Institut d'Electronique et de Télécommunications de Rennes, CNRS UMR 6164, Université de Rennes 1, 35000 Rennes, France

* Correspondence to philipp.delhougne@gmail.com.

**Achieving the very special condition of perfect absorption (PA) in a complex scattering enclosure promises to enable a wealth of applications in secure communication, precision sensing, wireless power transfer, analog signal processing and random lasing. Consequently, a lot of recent research effort was dedicated to proposing wave-front shaping protocols to implement coherent PA in a complex scattering environment. Here, instead of shaping the impinging wave front and tuning the system's attenuation, we propose the conceptually different route of solely tweaking the randomness of the complex scattering environment in order to achieve PA. We provide an experimental proof-of-concept in the microwave domain where we tune the randomness of a three-dimensional chaotic cavity with a programmable metasurface. We systematically investigate the achievability and extreme sensitivity of the PA condition. Our technique can impose a PA condition at over hundred distinct frequencies within a small frequency band, which allows us to propose and experimentally demonstrate a concrete practical application: receiver-powered secure wireless communication in a complex scattering enclosure. Our fundamentally new perspective on PA is applicable to all types of wave phenomena and our results foreshadow the large potential of this novel tool for minute wave control in sensing, communication and energy transfer.**

# Introduction

Perfect absorption (PA) of waves only occurs under very special conditions. Usually, an impedance mismatch between two media causes a partial reflection of wave energy at the interface[1]. The PA condition requires vanishing outgoing fields which can be associated with a zero eigenvalue of the non-Hermitian scattering matrix on the real frequency axis[2]. Various research tracks have provided routes toward PA of single-channel monochromatic excitation with a carefully engineered structure. On the one hand, PA of incident plane waves has been demonstrated with Salisbury screens and metasurfaces[3–8]. On the other hand, PA of a guided mode via critical coupling to a single-mode resonator has been reported[9,10]. For critical coupling, the coupling rate of the excitation channel must equal the resonator's intrinsic decay rate. More recently, these ideas were generalized to multi-channel monochromatic excitations. In these cases, the coherence of the incident channels becomes crucial: PA only occurs if the impinging wave front is an eigenvector of the scattering matrix that is associated with a zero eigenvalue[11,12]. Various experimental demonstrations with very regularly shaped planar and guided-mode structures were reported[13–17]. The extreme sensitivity to changes in the excitation or absorbing system makes PA a fascinating topic for fundamental research and is at the heart of its technological relevance in wave filtering, sensing, communication, energy harvesting and thermal emission control[12].

All of the above-described approaches rely on a carefully engineered material to achieve PA. Yet, given the ubiquity of disordered matter presenting random scattering effects for all wave phenomena, achieving PA in a random medium would unlock a wealth of new applications. Although in principle no fundamental limit prevents this, in practice it appears extremely complicated to adjust excitation and decay rate of a randomly scattering medium. Very recently, coherent PA in random scattering media has been studied[18–20] and demonstrated in proof-of-principle experiments[21,22]. These experiments consider systems in which attenuation is dominated by a tunable local loss center; a wide parameter space of frequencies and loss values is then searched to identify a setting in which a scattering matrix eigenvalue lies on the real frequency axis.

Here, we propose a novel route to implementing PA in a random scattering medium that does not rely on shaping the impinging wave front nor on tuning the system's attenuation. Instead, we judiciously tweak the medium's randomness which adjusts the coupling rate of the fixed excitation channel to the medium's modes. If losses are localized rather than global, this technique also enables adjustments of the medium's intrinsic decay rate by controlling the wave's exposure to loss centers. To demonstrate the feasibility of our concept, we consider a three-dimensional complex multi-mode scattering enclosure excited by a single guided mode in the microwave domain. The randomness of our medium, a chaotic cavity[23] with global rather than local absorption effects, is tuned with a programmable metasurface[24] that locally reconfigures the cavity's boundary conditions[25]. We thereby achieve PA irrespective of the impinging wave's amplitude and phase, without the need to control the amount of loss in the system, and moreover, we can dynamically switch PA on and off at multiple nearby frequencies. We systematically investigate the achievability and sensitivity of the PA condition with our technique. Then, for the first time, we demonstrate a practical application of PA in a complex scattering enclosure: we propose and implement a unique wireless communication protocol that is physically secure and receiver-powered. Our PA technique can also be implemented for any other type of wave phenomenon, for instance, using piezoelectric or acousto-optical modulators to tweak the randomness of a multimode fiber at optical frequencies[26,27].

## Results

**Principle and Achievability of PA.** The reflected power $\frac{P_{out}}{P_{in}}$ in a complex scattering enclosure such as a chaotic cavity is a statistically distributed quantity[28–30], such that sufficiently large changes of frequency or configuration can be interpreted as drawing a new value from the distribution. Thus, one may "accidentally" come across low reflectance values. The probability of "accidentally" finding an extreme value of this distribution, as is PA with zero rather than "low" reflectance, is incredibly low. In our complex scattering enclosure (an electrically large irregularly shaped metallic structure) depicted in **Figure 1a**, the probability to find $\frac{P_{out}}{P_{in}} \leq 2.5 \times 10^{-7}$ is $7.5 \times 10^{-6}$. To visualize the "achievability" of a given reflectance value, we plot the experimentally determined cumulative distribution function (CDF) $\Phi(\frac{P_{out}}{P_{in}})$ for our system in **Figure 1d**. $2.5 \times 10^{-7}$ is the lowest reflectance we can measure directly due to the presence of measurement noise. It is hence clear that "accidentally" coming across a PA condition is virtually impossible in our system.

At the core of our technique is the idea to drastically increase the probability of finding a PA by judiciously tweaking the randomness of the complex scattering enclosure. Our system is equipped with a 4x4 1-bit-programmable metasurface, as shown in **Figure 1a**. As detailed in **Methods** and **Supplementary Note 1**, each programmable meta-atom has two digitalized states, i.e., "0" and "1", corresponding to two opposite electromagnetic (EM) responses. Although the detailed EM response characteristics of the two states are irrelevant for our scheme as long as they are distinct, the idea behind our metasurface design is to emulate Dirichlet or Neuman boundary conditions via a phase difference of roughly $\pi$.[31,32] With our 16 reconfigurable meta-atoms placed inside the chaotic cavity, we can thus tweak the chaotic cavity's randomness in $2^{16}$ distinct ways. For a given target frequency $f_0$, we optimize the metasurface configuration in order to minimize the reflected power at $f_0$. An example is shown for $f_0 = 20.18$ GHz in **Figure 1b**. For reference, besides the optimal PA configuration, the reflected power spectrum for two random configurations of the metasurface is shown. While the PA configuration achieves $\frac{P_{out}}{P_{in}} = 2.28 \times 10^{-7}$, the reflected power is on the order of $6 \times 10^{-2}$ for the two random configurations. For the PA configuration, the corresponding trajectory of the frequency-dependent complex-valued reflection coefficient $S_{11}$ in the Argand diagram crosses the origin at $f_0$; for the random configurations, the trajectory does not get close to the origin.

Based on an evaluation of the lowest achievable reflectance at each of the 3201

considered frequency points in the $19 - 24 \text{ GHz}$ band, we evaluate the CDF of the lowest achievable $\frac{P_{out}}{P_{in}}$ values in our metasurface-programmable system. The corresponding curve (red line) in **Figure 1d** shows that the tuning mechanism drastically improves the probability to find $\frac{P_{out}}{P_{in}} \leq 2.5 \times 10^{-7}$ by four orders of magnitude from $7.5 \times 10^{-6}$ to $5.3 \times 10^{-2}$. With our metasurface-programmable system, the probability of finding a PA condition is thus tangible and indeed we encounter more than 100 frequencies at which PA is achievable within the considered frequency band. The area between the red and the blue curves in **Figure 1d** testifies to the substantial change of the CDF that we achieve.

To understand why we cannot achieve PA at any desired target frequency in our system, in **Figure 1e** we take a closer look in the Argand diagram at the cloud of accessible $S_{11}$ values at three representative frequencies. Ideally, such as in the first case, the cloud is centered on the origin. If the cloud only touches the origin as in the second case, the number of configurations yielding an $S_{11}$ close to the origin is much smaller. In the third case, if the cloud is far from the origin, achieving PA is impossible. Consequently, several possibilities to further enhance the achievability of PA clearly emerge. First, multi-bit-programmable meta-atoms would allow us to navigate a given $S_{11}$ dot in the Argand diagram to place it exactly on the origin, at least for all target frequencies for which the origin is covered by the cloud. The tunability mechanism of our meta-atoms being a varactor rather than a PIN diode, in principle it would only take a more elaborate biasing circuitry to enable continuous control of the EM response of each meta-atom. We therefore faithfully expect that continuous tuning of the meta-atoms would enable a further significant improvement of the CDF. Only in cases like the third in **Figure 1e** PA would likely continue to be inaccessible despite multi-bit programmability. This issue could be resolved by using more programmable meta-atoms which would increase the radius of the clouds. In **Figure 1d** we illustrate the reverse effect, namely of using fewer meta-atoms, on the PA achievability. With 8 instead of 16 1-bit programmable meta-atoms, the probability of $\frac{P_{out}}{P_{in}} \leq 2.5 \times 10^{-7}$ drops from $5.3 \times 10^{-2}$ to $7.0 \times 10^{-4}$ but still remains significantly above that of $2.5 \times 10^{-7}$ if the randomness is not tuned. For the cases of using only 4 or 2 programmable meta-atoms, it is evident in **Figure 1d** that the CDF curves are not smooth at low values of $\frac{P_{out}}{P_{in}}$ which indicates insufficient statistics (despite having run 45,000 realizations). In **Supplementary Note 4**, we study the impact of the complex scattering enclosure's $Q$-factor on the achievability of the PA condition and find that, counterintuitively, systems with lower absorption (i.e. higher $Q$) have a more favorable CDF for PA. Having outlined a clear route toward achieving PA at virtually any desired frequency, we point out that in fact achieving PA at a few frequencies as in our experiment is largely sufficient for many applications. Moreover, it is worthwhile acknowledging that wave-front-shaping-based coherent PA schemes as proposed in Refs.[21,22] rely on the frequency as free parameter to find a coherent PA condition, that is, by construction these methods are unable to achieve PA at any desired target frequency.

**Sensitivity.** The very narrow reflectance dip in **Figure 1b** already hints at the extreme sensitivity of the PA condition. Since this sensitivity is crucial for the concept's technological relevance, here we systematically study the impact of geometry and frequency detuning on the PA condition. A convenient way to implement geometry detuning in our setup is to tune the configuration of one meta-atom away from that imposed by the optimal metasurface configuration. Given the meta-atom's 1-bit-programmability in our setup, this geometric detuning is thus not continuous. For the 168 frequencies for which an optimal metasurface configuration achieves $\frac{P_{out}}{P_{in}} < 10^{-6}$ in our system, we measure the reflected power if one meta-atom is detuned. The resulting change of $\frac{P_{out}}{P_{in}}$ on a logarithmic scale is shown in **Figure 2a**, and the probability density function (PDF) $\theta$ of this quantity is plotted in **Figure 2b**. Except for the first two meta-atoms, detuning a single meta-atom is seen to systematically increase the reflected power by 40 to 50 dB, such that the detuned system with $\frac{P_{out}}{P_{in}}$ on the order of $10^{-2}$ does not present PA anymore. The PDF for the first two meta-atoms has two peaks, one around 40 to 50 dB and one around 0 dB. In roughly half of the considered PA cases, detuning one of the first two meta-atoms does therefore not destroy PA. Although the first two meta-atoms are the most distant ones from the port, since the other meta-atoms do not display any similar effect, we speculate that this effect must be an artefact due to some technical defect.

To investigate frequency detuning, first, we interpolate our measured reflection spectra. While interpolating the reflection magnitude is not possible due to its rapid variation (the narrowness of the dip), real and imaginary part of $S_{11}$ vary much slower (see **Figure 1c**) such that they can be interpolated faithfully. In **Figure 2c** we plot the change of $\frac{P_{out}}{P_{in}}$ on a logarithmic scale as the frequency is tuned above or below $f_0$. Occasionally, for very small detuning strengths, the reflected power decreases at first in one direction, as for the green line in the inset of **Figure 2c**. Due to the relatively coarse frequency sampling of our initial measurement, in these cases $f_0$ did not correspond precisely to the reflection dip's minimum. The high sensitivity of PA to frequency detuning is evidenced by the PDF of the change of $\frac{P_{out}}{P_{in}}$ on a logarithmic scale in **Figure 2d**. Upon detuning the frequency by a few MHz, the reflected power increases by 50 dB on average such that the PA condition is destroyed. The frequency detuning strength needed to destroy PA is thus at least an order of magnitude below the spectrum's characteristic correlation frequency $f_0/Q$ (see **Methods**) and at least four orders of magnitude smaller than $f_0$. Although for a given PA condition the reflection dip is not symmetrical for frequencies below and above $f_0$, the PDF is symmetrical.

We have studied two important sensitivities of PA related to geometry and frequency detuning. We faithfully expect that further extreme sensitivities exist, e.g. with respect to the spatial position of the port. An experimental investigation thereof is impossible since moving the port would simultaneously distort the geometry of our three-dimensional system. Nonetheless, since space and time tend to display analogous behavior in complex scattering environments, for instance, with respect to focusing[33], a very high sensitivity to spatial detuning is likely. Preliminary numerical results also point in that direction, as does

Ref.[21] for wave-front-shaping based coherent PA.

**Receiver-Powered Secure Wireless Communication.** We now propose and demonstrate a concrete practical application of our PA technique which leverages the two above-established features of (i) extreme sensitivity and (ii) achievability at many distinct nearby frequencies. The security of wireless communication is a major concern in many areas and to date most security features are software-based, i.e. using some sort of encryption of the transferred data. Here, we propose and demonstrate how the extreme sensitivity of the PA condition can be leveraged to implement secure communication on the hardware level, without any need for data encryption. A further striking feature of our proposed communication system is that it is receiver-powered such that the transmitter can transfer information at high security without any noteworthy power consumption. To establish a wireless communication channel between Alice and Bob, conventionally, Alice actively emits electromagnetic waves that are received by Bob. With the advent of programmable metasurfaces, engineering the communication channels by tuning the propagation environment has become conceivable[34,35]. However, programmable metasurfaces can also be used to imagine novel backscatter-communication concepts for information transfer that do not rely on an active generation of waves by Alice. For instance, Alice can instead communicate with Bob by using a programmable metasurface to focus already existing stray ambient waves on Bob's receiver.[36] Here, we propose a scheme whereby Alice configures the propagation environment to switch PA on and off at Bob's port. One potential area for deployment is future radiofrequency-based chip-to-chip communication[37].

If an eavesdropper Eve appears once the communication between Alice and Bob has been established, Eve's presence will detune the geometry and destroy PA, revealing her presence. We can thus focus on the case of Eve being part of the communication system since the beginning. If Alice alternates between the special PA configuration and a random one, or between two PA configurations for two distinct frequencies, she can emulate amplitude-shift keying or frequency-shift keying protocols. Although PA is only observable on Bob's port, with a high dynamic range Eve could detect the switching between two fixed configurations and decode the binary message up to a confusion regarding which configuration is "0" and which is "1". Since, however, we can easily achieve PA at several frequencies, "1" (resp. "0") can be defined as creating a PA (resp. nonPA) condition for Bob's port at a randomly chosen frequency. In **Figure 3a**, we show a selection of eight PA and eight nonPA configurations for our modified experimental setup including a second eavesdropper port as depicted in **Figure 3b**. Note that the ability to achieve eight PA configurations within a 0.04 GHz bandwidth is a unique feature of our PA technique based on tuned randomness, setting it clearly apart from wave-front shaping based approaches[21,22] (see also Discussion below). As illustrated in **Figure 3c**, for each bit, Alice randomly picks one of the eight configurations that yield PA (resp. nonPA) at Bob's port in order to send a "1" (resp. "0"). Eve can receive the power emitted by Bob as Bob measures his port's reflectance, or Eve can monitor the reflectance on her own port. In both cases, she will only observe a series of random signals irrespective of her dynamic range. In **Figure 1c** we demonstrate with *in situ* measurements that our proposed communication scheme flawlessly transfers information from Alice to Bob at high security: Bob correctly receives all bits and Eve cannot extract any information about what bits Alice sends to Bob. Finally, imagine the presence of a manipulator Mike, also equipped with a programmable metasurface. Mike could not pretend to be Alice and impose PA on Bob's port without Alice's collaboration.

## Discussion and Conclusion

The extremely narrow reflection dip associated with the PA condition, as seen in **Figure 1b**, makes our experimental setup very attractive for wave filtering applications. Usually, filter applications require elaborate fabrication techniques to achieve very high $Q$-factors. In contrast, our complex scattering enclosure is made from simple copper tape and its average $Q$-factor is only on the order of $10^3$ (see **Methods**). Nonetheless, the PA condition alters the resonance width distribution[38,39] and imposes an extremely narrow linewidth on the mode that enables PA (i.e. the mode that lies on the real frequency axis). In **Supplementary Note 5**, we outline further potential applications of our PA technique to nondestructive high-precision evaluation and long-range efficient energy transfer. Here, we now comment on a few advantages of our method to tune the random medium's disorder over the methods reported in Refs.[21,22] which rely on shaping the impinging wave front and simultaneously tuning the random medium's level of attenuation. First, from an experimental point of view, the simplicity of our setup (Arduino microcontroller and printed-circuit-board metasurface) is in sharp contrast to the immense hardware cost entailed by the need for (i) precise amplitude and phase tuning of the incident wave front, e.g. with 16 IQ-modulators in Ref.[21], and (ii) *in situ* control of the system's attenuation (typically via a single tunable loss center). Moreover, many realistic scattering systems like our 3D enclosure present significant homogeneous attenuation which prevents full *in situ* control of attenuation with a single loss center as in Refs.[21,22]. Our proposal's ease of implementation (low hardware complexity and cost, no need to tune attenuation) will be a key enabler of the proliferation of PA-based applications in real life. Second, from a statistical point of view, since our technique allows us to access over 160 instances of PA conditions within the considered frequency band (and potentially significantly more with multi-bit programmable meta-atoms), we were able to study statistical properties such as $\Phi(\frac{P_{out}}{P_{in}})$ and $\theta(\Delta \frac{P_{out}}{P_{in}} [\mathrm{dB}])$. In contrast to wave-front-shaping schemes where the frequency is needed as free parameter such that only a single PA condition (within a rather large frequency interval) is identified, our approach thus enables a fuller study of PA in connection with its statistical characteristics. Our technique's unique ability to achieve PA conditions at multiple nearby frequencies also underpins our proposed protocol for secure communication.

Our technique of tuning the medium's randomness to achieve PA can be generalized to multiple excitation channels in different ways. One possibility is to engineer the disorder such that one eigenvalue of the scattering matrix is zero, enabling coherent PA at that frequency. Alternatively, one could force all diagonal scattering matrix entries to zero such that the outgoing field vanishes for any impinging wave front. The scattering matrix would then have multiple zero eigenvalues which results in degenerate coherent-PA modes[2,40]. One may also wonder if our work could straight-forwardly be extended to achieving the opposite of PA, namely perfect reflection at the interface. However, since homogeneous absorption in the scattering enclosure inevitably entails some loss that cannot be avoided by the waves, the reverberating waves incident on the port will never be able to create a perfect destructive interference with the waves leaving the port – unless gain-programmable meta-atom[41] inclusions are used[42]. Perfect reflection may have technological relevance, for instance, to prevent an intruder's mobile device from emitting any waves and hence from communicating. Moreover, the perfectly reflected channel would act like an electromagnetic sink[43] with interesting sub-diffraction features.

To summarize, we proposed and demonstrated how tuning a random medium's disorder can constitute a route to achieving the very special condition of PA without the need to control the impinging wave front and the level of attenuation in the medium. In our experiment which leverages a programmable metasurface to tune a complex scattering enclosure, we thoroughly investigated the achievability of PA with our approach as well as the extreme sensitivity of PA to geometry or frequency detuning. Building on these results, we went on to propose and demonstrate a protocol for receiver-powered secure wireless communication. We expect our technique to impact many areas in need of minute wave control at microwave frequencies as well as for other wave phenomena.

## Methods

**Design of programmable metasurface.** The programmable metasurface is an ultrathin planar array of electronically reprogrammable meta-atoms. Programmable metasurfaces are a promising young member of the metamaterial family thanks to the ease of fabrication and their unique capability to manipulate electromagnetic fields in a reprogrammable manner. The designed programmable metasurface in our experiment consists of a 4x4 array of meta-atoms operating in the K-band around 21 GHz. The metasurface is based on Rogers 4003C substrate (1.5 mm thick, dielectric constant of 3.55 and loss tangent of 0.0027). Each meta-atom consists of a mushroom structure including a varactor (MACOM MAVR-011020-1411) which endows it with programmability. The detailed mushroom structure can be found in **Supplementary Note 1**. In our design, the metasurface is equipped with two shift registers and reprogrammed via an Arduino microcontroller. Although our varactor-based design in principle enables gray-scale tuning, our control circuitry restricts us to binary control (on/off). The two states of each meta-atom emulate perfect-electric-conductor and perfect-magnetic-conductor-like behavior at the center of the operation band, i.e. their response has a phase difference of roughly $\pi$. Note that the specific details of the meta-atom response are not important in our work; what matters is that reconfiguring the meta-atom has a notable impact on the cavity wave field. Further details on the metasurface are provided in **Supplementary Note 1**.

**Proof-of-principle system.** The complex scattering environment in our experiment is a chaotic cavity[23] of dimensions 11×11×5.5 cm³ with a quality factor of $Q = 1063$. The reflection spectrum's average correlation frequency is thus $\Delta f_{corr} = \frac{f_0}{Q} \approx 20$ MHz. One corner is deformed into a sphere octant to introduce wave chaos. Electrically large cavities of irregular shape are termed "wave-chaotic" since the separation of two rays launched from the same location in slightly different directions increases exponentially in time. A 8×8 cm² area on one cavity wall is equipped with our home-made 4x4 programmable metasurface. A WR42 rectangular waveguide is used to pump electromagnetic energy into the cavity. A 1m-long 50-Ω coaxial cable connects the WR42 waveguide port to the VNA in order to measure the frequency-dependent reflection coefficient ($S_{11}$). To keep the measurement noise low, we set the VNA's intermediate-frequency bandwidth to 5 kHz and work with a power of 0 dBm. The wall opposite to the programmable metasurface can be perforated with holes to introduce additional loss channels in a controlled manner. Further details on the experimental setup are provided in **Supplementary Note 2**.

**Optimal metasurface configuration.** Since there is no analytical forward model to predict the reflection spectrum as a function of metasurface configuration, identifying the optimal configuration to approach PA as closely as possible at a desired target frequency is not straightforward. Most reports in the literature on tuning the randomness of a complex medium in other contexts relied therefore on lengthy sequential iterative trial-and-error methods implemented experimentally[25,34]. Such iterative solvers are not guaranteed to identify the globally optimal configuration and entail an immense experimental cost. Therefore, statistical analysis as in our manuscript is precluded. Instead, we measure the reflection spectrum for all possible $2^{16}$ metasurface configurations, such that we can identify the globally optimal configuration at any desired frequency using any desired subset of meta-atoms.

## Acknowledgments

This work was supported in part by the Air Force Office of Scientific Research (AFOSR) (No. FA9550-18-1-0187). P.d.H. was supported in part by the French "Agence Nationale de la Recherche" under reference ANR-17-ASTR-0017.

## Author Contributions

P.d.H. conceived the idea, analyzed the data and wrote the paper. M.F.I. and P.d.H. conducted the experimental work. All authors contributed with thorough discussions and commented on the manuscript.

## Additional Information

**Supplementary Information** accompanies this article.

**Competing Interests**: The authors declare no competing financial interests.

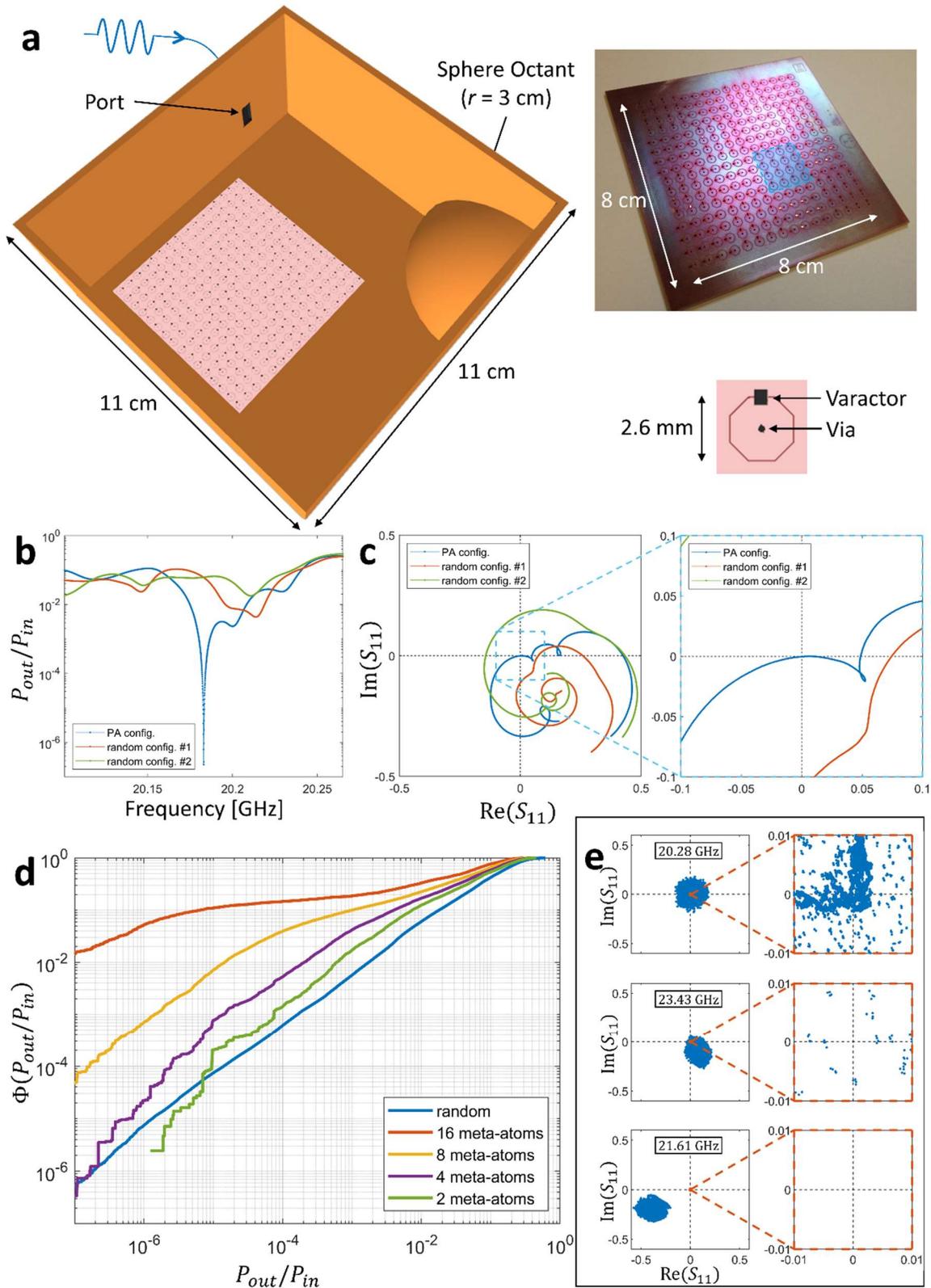

**Figure 1. Experimental setup, PA principle and PA achievability. a,** A monochromatic

guided mode excites a complex scattering enclosure (top wall removed to show interior), including a sphere octant to perturb the otherwise regular cavity modes, and our 4x4 programmable metasurface. One meta-atom is highlighted in the photographic image of the programmable metasurface. Each 1-bit-programmable meta-atom consists of 16 mushroom structures. Details of the latter are shown as inset. **b,** Experimentally measured reflected power for the PA metasurface configuration ($f_0 = 20.18$ GHz) as well as for two random metasurface configurations. **c,** Trajectories of the frequency-dependent complex-valued reflection coefficient $S_{11}$ in the Argand diagram corresponding to the three curves in **b**. A zoom is provided as inset. **d,** Experimentally determined cumulative distribution function $\Phi(P_{out}/P_{in})$ of the reflected power for our complex scattering enclosure. The blue curve is based on the $19 - 24$ GHz band and all possible metasurface configurations. The remaining lines consider for each frequency point only the reflected power corresponding to the optimal metasurface configuration, for different numbers of programmable meta-atoms. **e,** Clouds of all accessible reflection coefficients with our 1-bit programmable metasurface in the Argand diagram for three selected frequencies. A zoom is provided as inset.

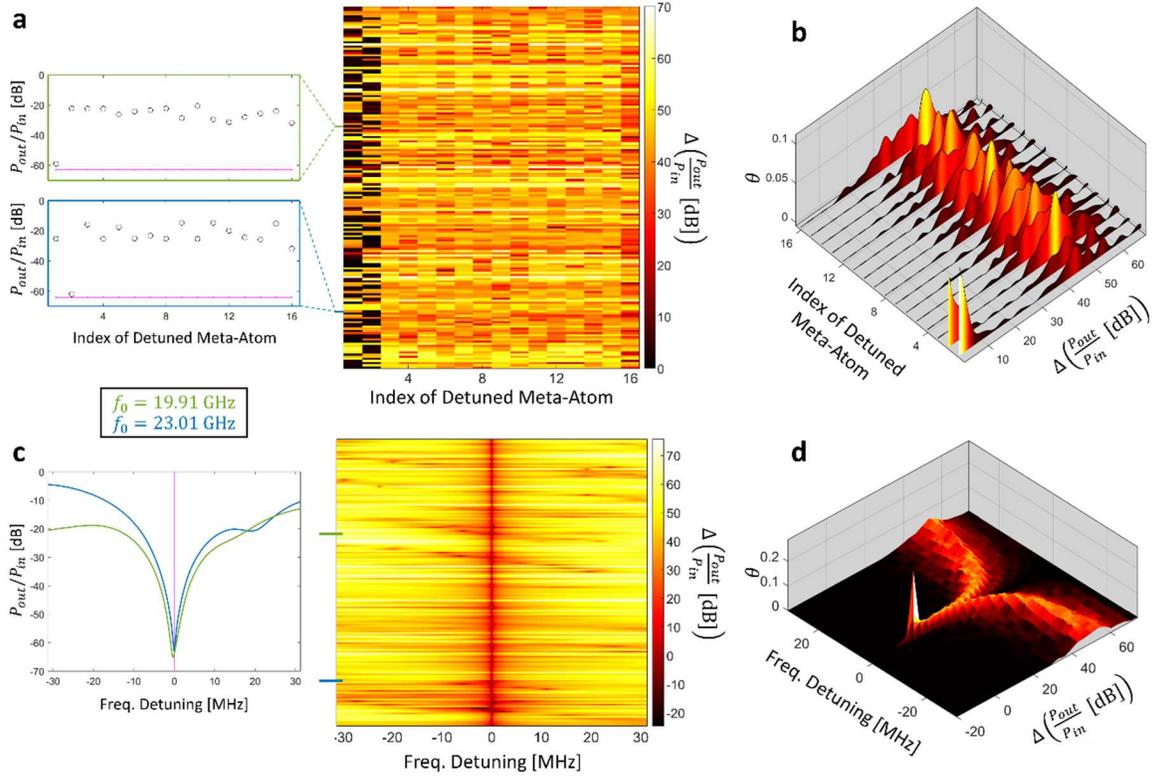

**Figure 2. Sensitivity of PA condition to geometry and frequency detuning. a,** Sensitivity to geometry detuning. The change of $\frac{P_{out}}{P_{in}}$ on a logarithmic scale upon flipping one meta-atom's configuration from "0" to "1" or vice versa is shown for 168 frequencies for which the optimal metasurface configuration achieves $\frac{P_{out}}{P_{in}} < 10^{-6}$. The inset shows for two representative frequencies the reflected power with the optimal metasurface configuration (purple line) as well as for the 16 configurations with one detuned meta-atom. **b,** For each meta-atom, the PDF $\theta$ of the change of $\frac{P_{out}}{P_{in}}$ on a logarithmic scale upon detuning is shown, evaluated based on the 168 frequencies considered in **a**. **c,** Sensitivity to frequency detuning. The change of $\frac{P_{out}}{P_{in}}$ on a logarithmic scale upon detuning the PA frequency $f_0$ is shown for 168 frequencies for which the optimal metasurface configuration achieves $\frac{P_{out}}{P_{in}} < 10^{-6}$. The inset presents the reflected power as function of the detuning frequency for two representative frequencies. **d,** For each considered frequency-detuning strength, the PDF $\theta$ of the change of $\frac{P_{out}}{P_{in}}$ on a logarithmic scale is shown, evaluated based on the 168 frequencies considered in **d**.

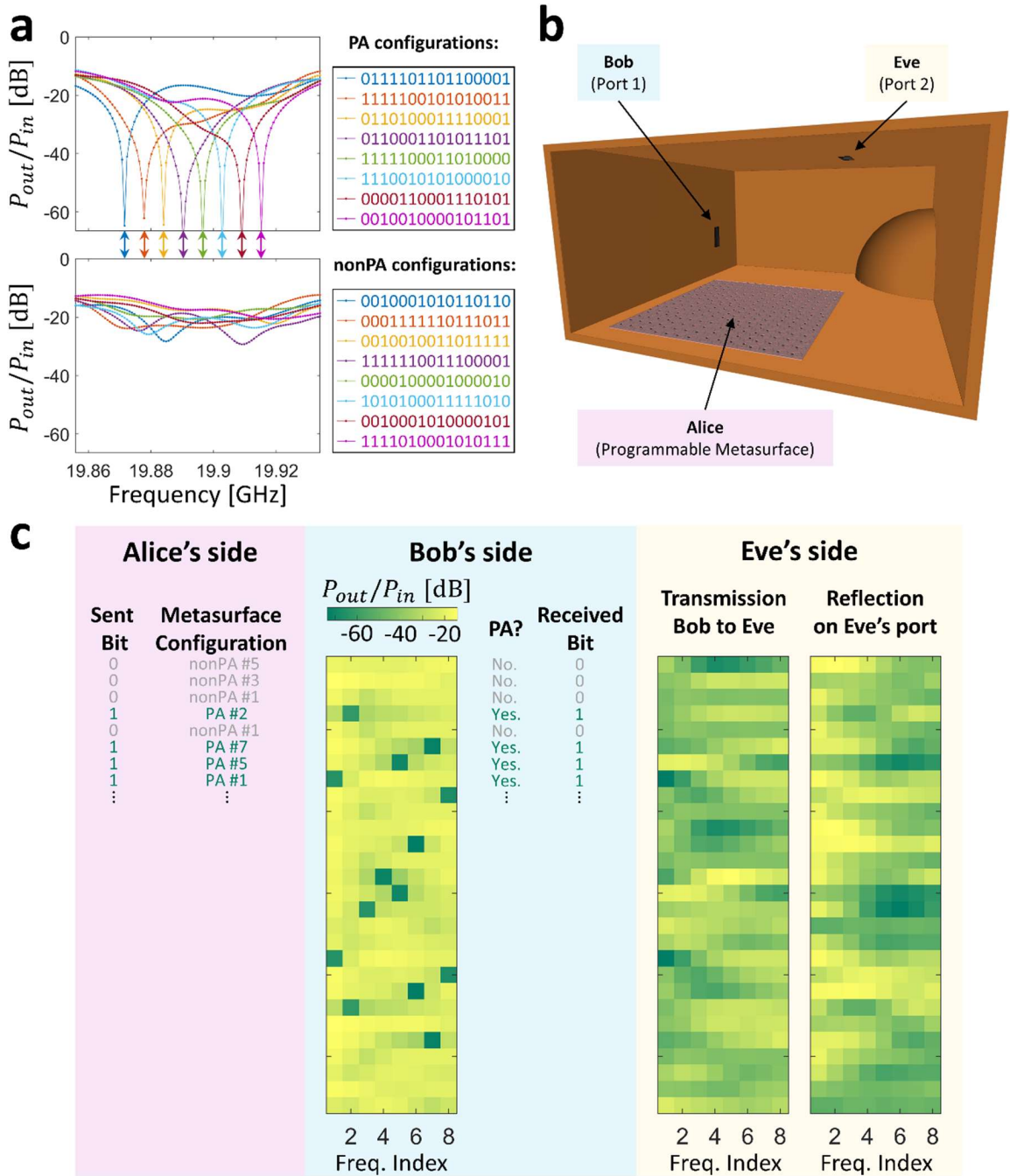

**Figure 3. Receiver-powered secure wireless communication. a,** Selection of eight metasurface configurations yielding PA at eight corresponding frequencies (indicated with arrows) as well as eight configurations not yielding PA at these frequencies. **b,** Experimental setup including a second (eavesdropper) port (front wall removed to show interior). **c,** *In situ* measurements of information transfer from Alice to Bob using randomly chosen PA (nonPA) configurations out of those identified in **a** to encode "1" ("0"). Simultaneously, the signals that Eve can obtain by receiving the signals emitted by Bob or measuring her port's reflectance are measured. Both are random and do not reveal what information Alice sends to Bob.

# Supplementary Materials for

# Perfect Absorption in a Metasurface-Programmable Complex Scattering Enclosure

Mohammadreza F. Imani[1], David R. Smith[1], Philipp del Hougne[2*]

[1] Department of Electrical and Computer Engineering, Duke University, Durham, North Carolina 27708, USA

[2] Institut d'Electronique et de Télécommunications de Rennes, CNRS UMR 6164, Université de Rennes 1, 35000 Rennes, France

* Correspondence to philipp.delhougne@gmail.com.

## Supplementary Note 1. Programmable Metasurface Design

The literature contains by now a number of reports on designs for programmable metasurfaces[1]. The latter are also referred to as "binary tunable metasurface"[2], "spatial microwave modulator"[3], "tunable impedance surface"[4,5] or "reconfigurable intelligent surface"[6]. We stress that the detailed mechanism that allows the metasurface to tune the random medium's disorder is irrelevant for the concept of PA by disorder engineering that we introduce in our manuscript. In fact, the generality of the concept of PA by disorder engineering extends even beyond the realm of programmable metasurfaces. Other tuning mechanisms could be used to tweak the randomness. For instance, in the optical domain, a common random medium is the multimode fiber whose randomness can be tuned with piezoelectric or acousto-optical modulators[7,8].

For our experiment, first of all, it is important to note that a chaotic wave field does not possess a dominant field polarization. Unlike free-space applications, our work in a complex scattering enclosure does not need the programmable meta-atoms to impose specific phase or amplitude changes upon reconfiguration. What matters is that the field is notably impacted by reconfiguring the meta-atoms. From a fundamental point of view, reconfiguring the meta-atoms alters the cavity's boundary conditions which results in a perturbation of the cavity eigenmodes.

The design of our programmable metasurface shown in **Supplementary Figure 1** is inspired by early works on high-impedance electromagnetic surfaces[4,5]. We use mushroom structures for which an octagon is inscribed in a circle of 2 mm diameter. A 0.5-mm-diameter via is connected to the ground plane. A MACOM MAVR-011020-1411 varactor is used. Simulations in a full-wave electromagnetic solver (CST Microwave Studio) show that under normal plane wave illumination the amplitude of a wave reflected off a mushroom unit cell remains close to unity across the considered frequency interval for both considered dc bias voltages of the varactor (0V and 5V). At the same time, the phase of the two states changes by roughly $\pi$. Due to component and fabrication tolerances we find that the metasurface has its largest impact on the wave field at slightly higher frequencies (19.5 – 23.5 GHz) than suggested by these simulations.

We group a 4x4 array of mushroom unit cells together and control them with the same bias voltage. This constitutes one meta-atom. Our 16 meta-atoms are then assembled in a 4x4 array. The varactors of the mushroom unit cells are rotated by 90° between neighboring meta-atoms to ensure that both independent field polarizations can be tuned.

Finally, we discuss the programmability constraint of our meta-atoms. In our current setup, we can only bias the varactor with two dc voltages (0V or 5V) such that we are restricted to 1-bit programmability. This has been identified as a major limit on the performance in the main text. With multi-bit programmability, it would be possible to achieve PA at least for every frequency for which the cloud of $S_{11}$ values covers the origin (see **Figure 1e** of the main text). While many metasurfaces in the literature are designed exclusively for 1-bit operation, e.g. because PIN diodes are used, our metasurface uses varactors and could in principle be biased with continuous dc voltages to provoke a continuously tunable response. In our setup, it is only the control circuitry that currently prevents us from using multi-bit programmability. Continuous biasing of the meta-atoms could be implemented, for instance, with digital-to-analog converters (DAC).

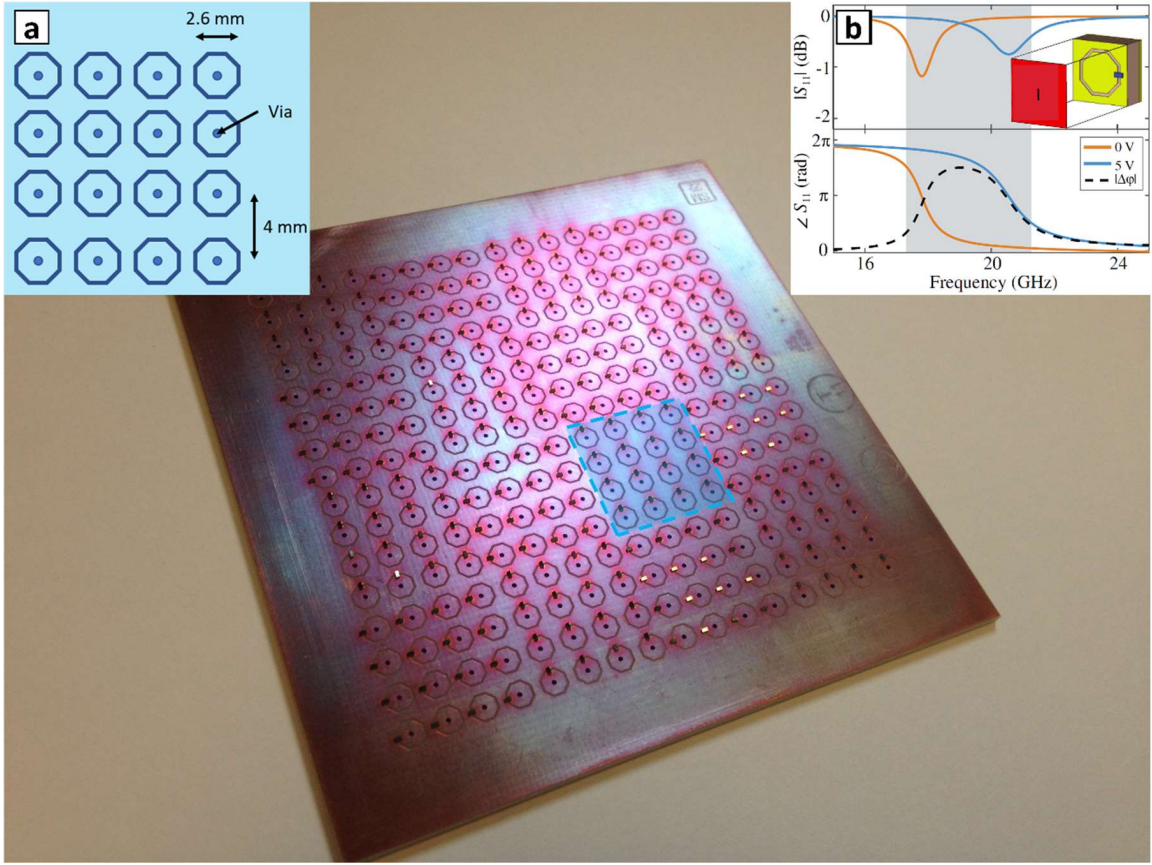

**Supplementary Figure 1. Programmable metasurface design.** Photographic image of our 4x4 programmable metasurface (photo: courtesy of Dr. Timothy Sleasman). One meta-atom consisting of 4x4 mushroom unit cells is highlighted. **a**, Geometrical details of the meta-atom. **b**, Response of mushroom unit cell under normal plane wave illumination obtained in simulation[5].

# Supplementary Note 2. Experimental Settings

In **Supplementary Figure 2** we present a global schematic overview of our measurement setup. One cavity wall has been removed in the drawing to show the inside including the deformity, the 4x4 programmable metasurface and the WR42 port. The control circuitry to program the metasurface involves a computer, an Arduino microcontroller and two shift registers. The WR42 port is connected to a port of the vector network analyzer (VNA) via a 1m-long semi-rigid 50-Ω coaxial cable. The VNA pumps electromagnetic energy into the cavity and monitors the reflected power. The cavity's $Q$-factor can be altered by adding slots to the cavity wall as seen in **Supplementary Figure 3**.

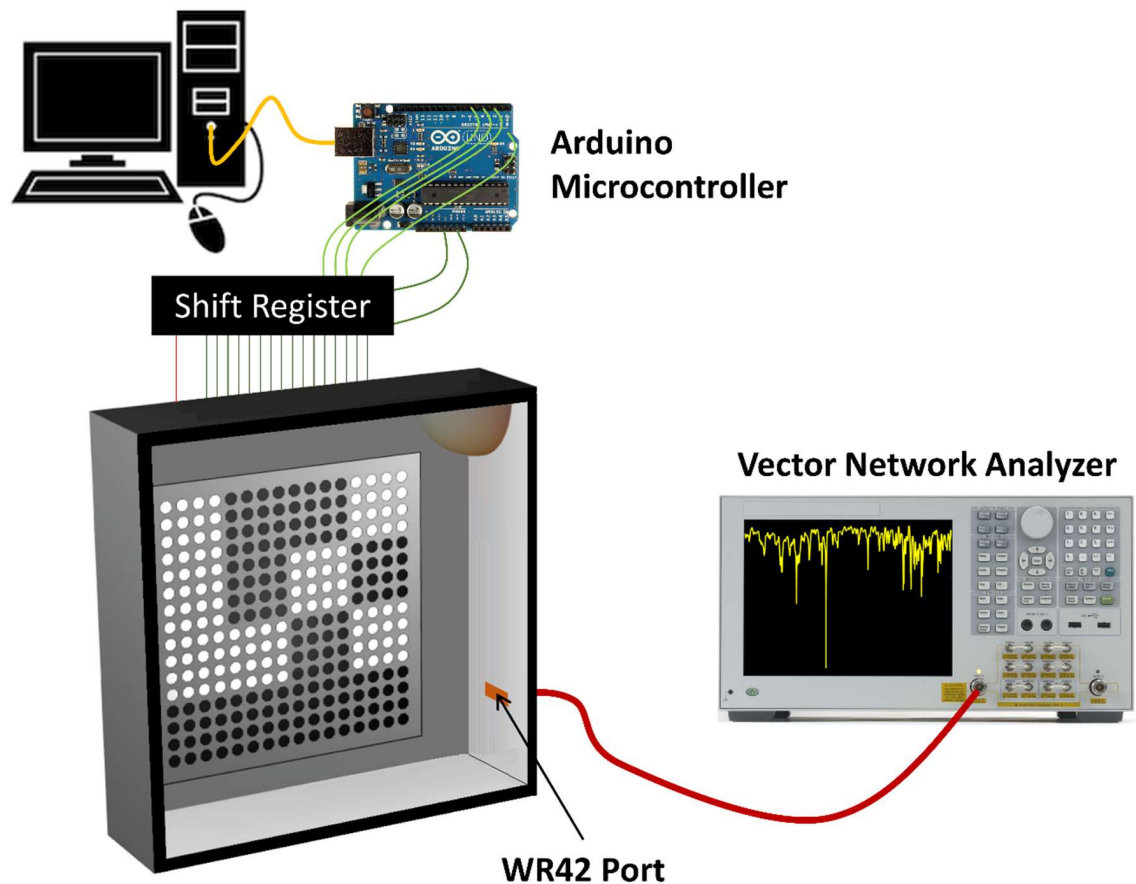

**Supplementary Figure 2. Proof-of-principle experimental setup.** The schematic drawing illustrates how the cavity port is connected to a VNA to measure the reflection spectrum and how a computer programs the metasurface via Arduino and shift register. One cavity wall (the one containing a variable number of slots to control $Q$) has been removed to show the inside of the cavity.

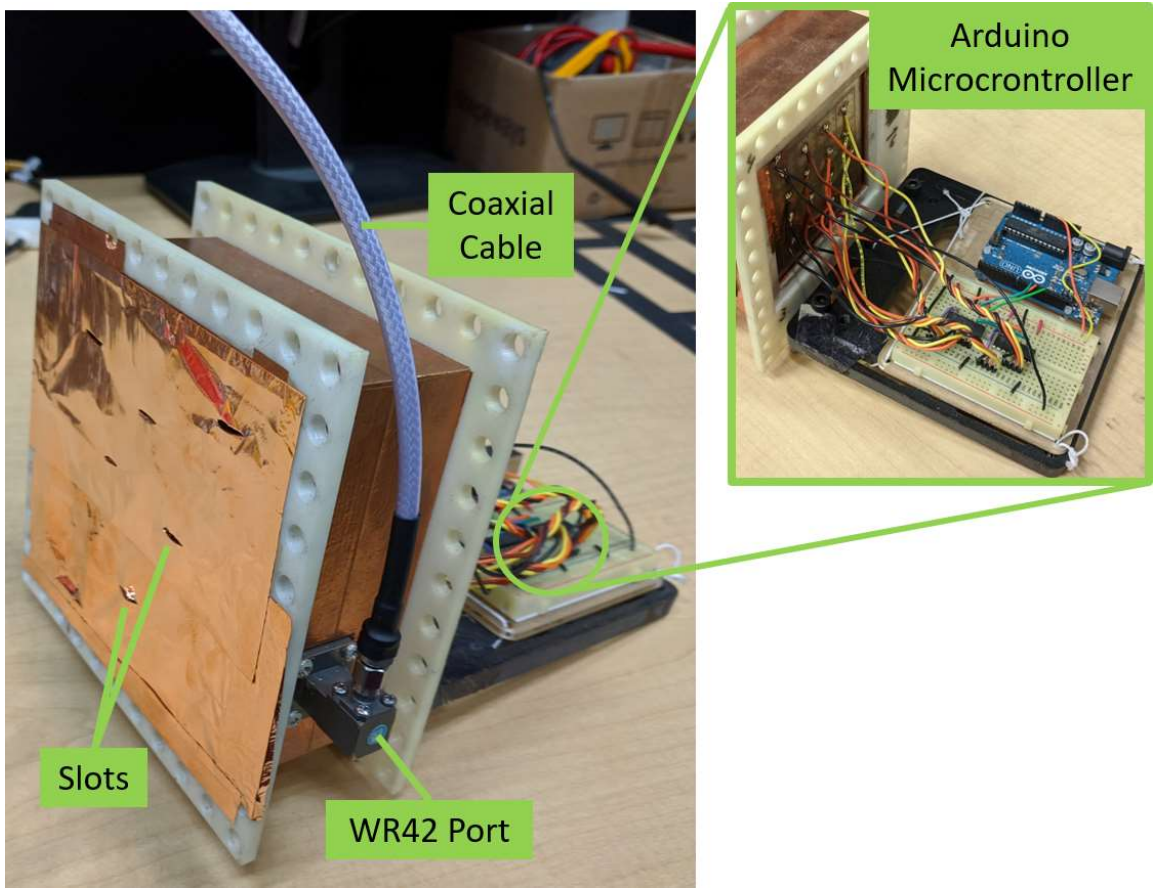

**Supplementary Figure 3. Photographic images of experimental setup.** The main picture shows our chaotic cavity with the front wall containing a few slots to adjust the cavity's quality factor and the WR42 port to pump electromagnetic energy from the VNA into the cavity to measure the reflectance. A semi-rigid coaxial cable connects the port to the VNA. The inset shows the control circuitry used to program the metasurface consisting of an Arduino microcontroller and two shift registers.

# Supplementary Note 3. Discussion of Metasurface Absorption

Here, we clarify whether the metasurface configuration acts in some way as local absorption mechanism. In **Supplementary Figure 4** we evaluate for each frequency the percentage $M$ of meta-atoms in the "on" state (5V varactor bias) for the metasurface configuration that minimizes the reflected power. $M$ is not predominantly close to zero or unity, and its frequency-average of 0.49 indicates that there is no preference for either state. Moreover, given that the reflection-suppression performance is highly dependent on the metasurface configuration, we can also exclude that it is the bare presence of the metasurface (irrespective of its configuration) that enables our results. Hence, we conclude that the metasurface certainly contributes to the global absorption in the cavity, as do all the other cavity walls, but that the role of its configuration is primarily to engineer the controllable component of the Green's function, $\bar{\bar{G}}_{s,c}$ (see **Supplementary Note 6**).

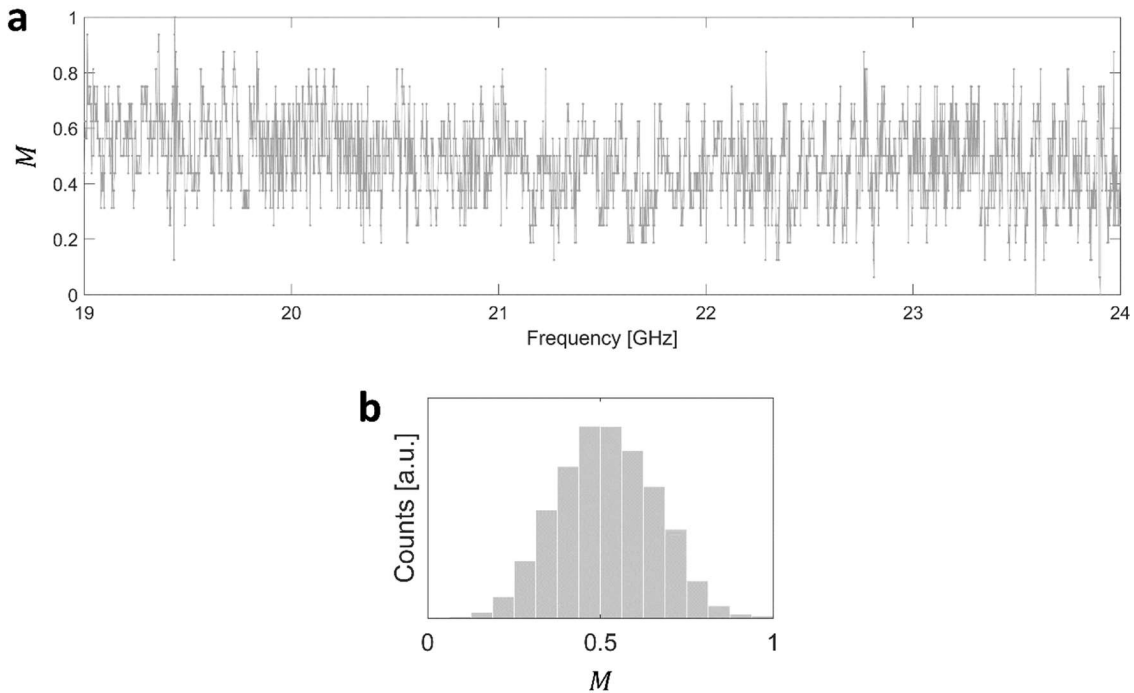

**Supplementary Figure 4. Percentage of meta-atoms in "on" state for optimal configuration. a,** Percentage $M$ of meta-atoms in the "on" state for the optimal configuration at each frequency. **b,** Histogram of the values of $M$ seen in **a**.

# Supplementary Note 4. Impact of $Q$-factor on Achievability of PA condition

In the main text, we explored the impact of the number of programmable meta-atoms on the achievability of the PA condition in **Figure 1d**. Here, we perform a similar analysis to investigate the impact of the complex scattering enclosure's $Q$-factor on the achievability of PA. To lower the system's $Q$-factor, we add slots to the wall opposite the metasurface (see **Supplementary Figure 3**). The more slots there are and the larger they are, the more energy leaks out of the cavity, resulting in a lower $Q$. $Q$ is estimated as $\pi f_0/\mu$, where $\mu$ is the exponential decay constant of the impulse response (obtained via an inverse Fourier transform of $S_{11}$) averaged over metasurface configurations.

In **Supplementary Figure 5**, we reproduce the results for the system considered in the main text in blue and add two cases corresponding to progressively lower $Q$-factors in red and yellow. Note that the latter two are based on the same $19-24$ GHz band but only use 801 frequency points; this explains why for low values of $\frac{P_{out}}{P_{in}}$ the red and yellow curves are not smooth due to insufficient statistics. Nonetheless, two conclusions can be drawn. First, the CDF for the random system does not significantly depend on $Q$. Second, the CDF of the tailored random system deteriorates slightly as $Q$ decreases. At first sight, this may appear counterintuitive: a system with more loss is worse at perfectly absorbing a wave. The reason for this, however, is that PA is based on the interplay of loss and interference. Indeed, according to Ref.[9] coherent PA can be implemented in principle with an arbitrarily small amount of loss. At the same time, however, the control over the wave field that is necessary to engineer the wave interferences inside the complex scattering enclosure increases with $Q$.

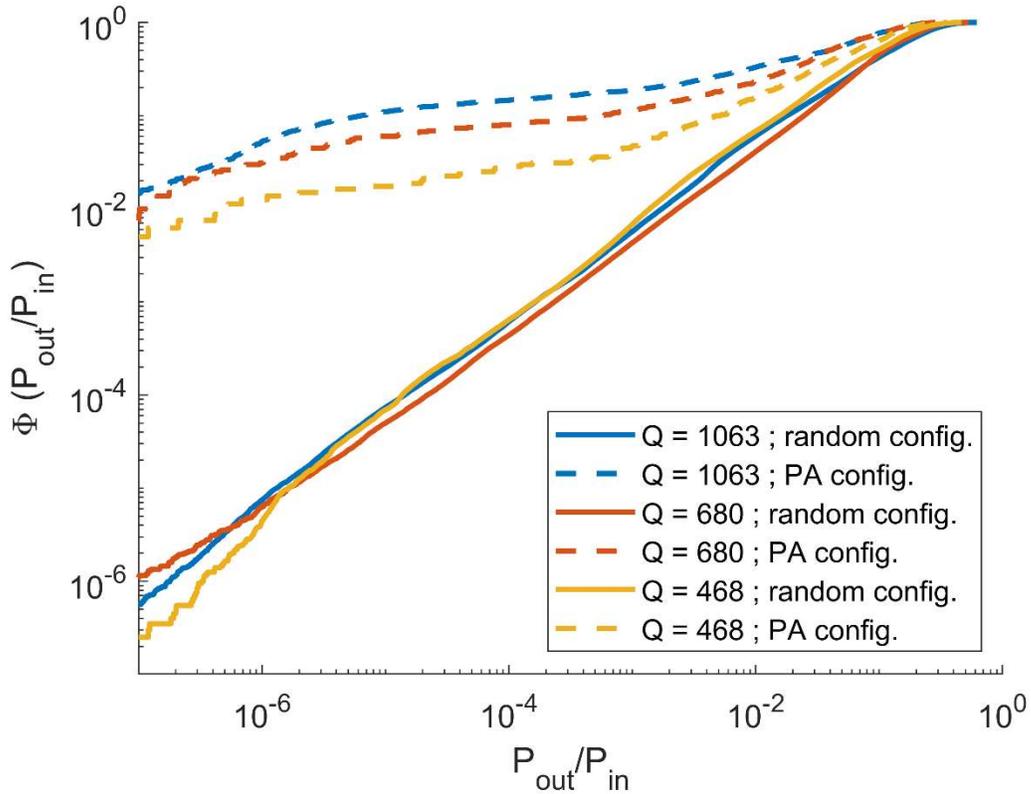

**Supplementary Figure 5. CDF of reflected power at various $Q$-factors.** The experimentally determined cumulative distribution function $\Phi(P_{out}/P_{in})$ for complex scattering enclosures with three different $Q$-factors (color-coded, see legend) are shown. The continuous curves are based on the $19-24$ GHz band and all possible metasurface configurations. The dashed lines consider for each frequency point only the reflected power corresponding to the optimal metasurface configuration (using all 16 programmable meta-atoms).

To quantify the impact of $Q$ on the control over the wave field, we evaluate the distance of the center of the cloud of possible $S_{11}$ values from the origin in the Argand diagram, as well as the cloud's size, and average these values over all measured frequencies. **Supplementary Figure 6** demonstrates that on average the distance of the cloud's center from the origin does not vary with $Q$ in an appreciable manner whereas the size of the cloud of accessible $S_{11}$ values is larger at higher $Q$. Therefore, at higher $Q$, it is more probable that the cloud covers the origin for a given frequency and that it contains a point that is extremely close to the origin. Thus, it is easier to achieve PA at higher $Q$.

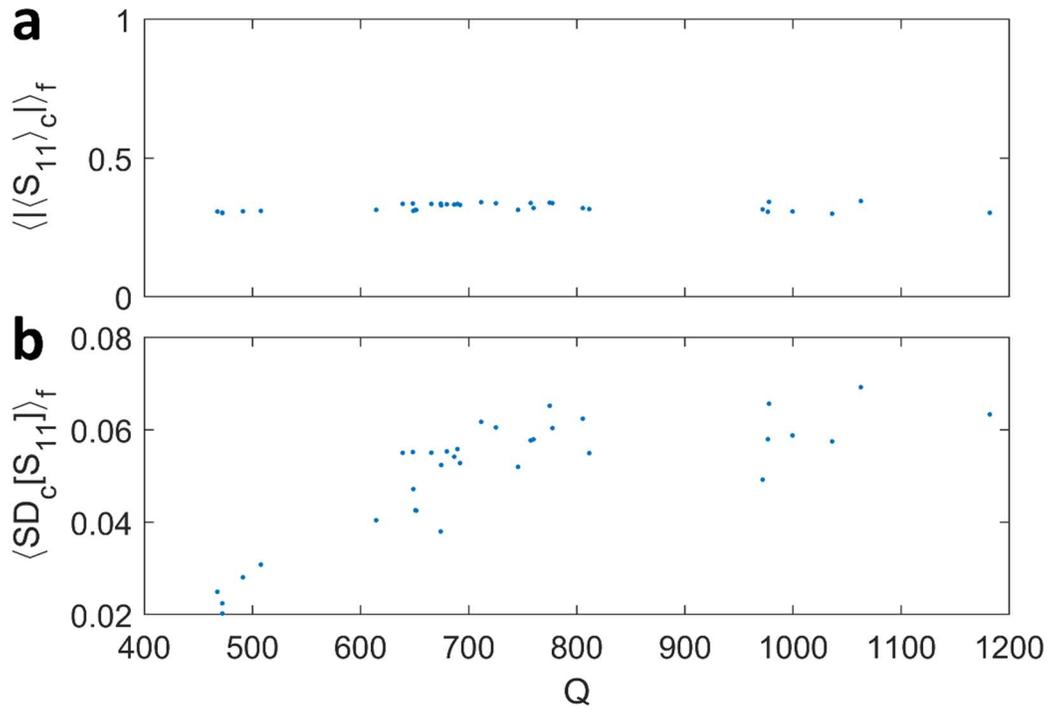

**Supplementary Figure 6.** Dependence on $Q$ of the frequency-averaged distance from the origin (a) and diameter (b) of the cloud of accessible $S_{11}$ values in the Argand diagram.

## Supplementary Note 5. Further Applications

**Nondestructive High-Precision Evaluation.** Rapid high-resolution non-destructive inspection of dielectric materials such as composites (e.g. glass-fiber-reinforced polymers) is of growing importance in several industrial sectors (automotive, aerospace, marine and structural engineering) to detect critical defects (disbonds, voids, delaminations, …) during manufacturing or in service that endanger the composite's structural integrity.[10] Current microwave far-field imaging techniques suffer from low resolution and time-reversal-based techniques require costly broadband hardware and sensor arrays[11]. Based on PA, the component under test could simply be placed in a metallic enclosure whose randomness has been configured such that PA at a specific frequency is achieved if the component is flawless. If the component is faulty, the system geometry is detuned from the calibrated PA condition and PA is not observed, such that the operator can conclude that a defect occurred. The advantages over other wave-chaos based monitoring schemes[12–14] are (i) the extreme sensitivity to tiny defects and (ii) the simplicity with which the observation can be made (no need for elaborate data processing, broadband operation, or multiple channels). A further application of the same principle is the distinction of liquids differing in concentration of a substance. In food, pharmaceutical and chemical industries it is important to monitor with high accuracy in real time that the concentration of various mixture solutions is constant at the desired level[15–17]. Permittivity and conductivity of the solution are concentration-dependent such that a wrong concentration detunes the PA condition and is easily and directly recognized. Beyond simple monitoring, the localization of objects inside the enclosure would be possible by learning a dictionary of unique metasurface configurations that yield PA at a probe port for different object positions[18].

**Efficient power transfer.** Wireless power transfer, energy harvesting and electronic warfare all necessitate the ability to focus waves on a lossy target with high efficiency, across large distances, and despite complex scattering environments. Current approaches like time-reversal focusing[19,20] rely on costly hardware (wideband multichannel wave control) and suffer from relatively low efficiency. Assuming that the target is the dominant loss mechanism in the system, recent theoretical[21,22] and experimental[23,24] studies showed that wave-front-shaping based coherent PA enables maximal absorption of the energy by the target. Our scheme enables blind single-channel focusing on an embedded absorbing target that dominates absorption in the system, without any need for adjusting phase or amplitude of the wave, nor for mechanical or electronic tuning of the target, nor for knowing the target's location. We speculate that if the target is equipped with a non-linear component (e.g. a rectifier circuit), the PA configuration of the metasurface could be identified without direct feedback from the reflected-power spectrum at the port, by monitoring the strength of non-linear harmonics re-emitted by the target.[25] Nonetheless, exploring realistic scenarios in which the focusing target is not the (only) dominating loss mechanism will require further attention in the PA literature, irrespective of whether wave-front shaping or tuning of the randomness is the utilized technique to achieve PA.

## Supplementary Note 6. Green's Function Interpretation

Here, we discuss some theoretical background of our proof-of-principle experiment. Flux conservation at the interface imposes $1 - |S_{11}|^2 = P_{abs} + P_{rad}$, where we have normalized the total incident power to unity. $P_{abs}$ denotes the power lost in the port due to Ohmic losses; the latter are typically very low, and more importantly, they do not depend on the metasurface configuration. $P_{rad}$ denotes the power radiated from the port into the enclosure and can be expressed in terms of the imaginary part of the Green's function[26]. Since the port is not point-like but wavelength-sized, at a given frequency we integrate the imaginary part of the dyadic Green's function $\bar{\bar{G}}(\mathbf{r}_1, \mathbf{r}_2)$ projected onto the port's aperture field $\mathbf{e}(\mathbf{r})$ over the port's aperture $\Omega$:[27,28]

$$P_{rad} \propto \int_\Omega \int_\Omega \mathbf{e}(\mathbf{r}_1) \cdot \mathrm{Im}\left[\bar{\bar{G}}(\mathbf{r}_1, \mathbf{r}_2)\right] \cdot \mathbf{e}^T(\mathbf{r}_2) \, d\mathbf{r}_1 d\mathbf{r}_2, \qquad (S1)$$

where $^T$ denotes the transpose and $\mathbf{r}_1, \mathbf{r}_2 \in \Omega$. To stress the role of the scattering environment, and in particular its tunable component, we first write $\bar{\bar{G}}$ as sum of the free-space Green's function and the scattering contribution, $\bar{\bar{G}} = \bar{\bar{G}}_0 + \bar{\bar{G}}_s$, and then further decompose $\bar{\bar{G}}_s$ into its average over all metasurface configurations and the contribution of one particular configuration:

$$\bar{\bar{G}} = \bar{\bar{G}}_0 + \langle \bar{\bar{G}}_s \rangle_c + \bar{\bar{G}}_{s,c}. \qquad (S2)$$

To achieve our goal of zero reflected power, the coupling rate of the excitation channel must equal the chaotic cavity's intrinsic decay rate. $\bar{\bar{G}}_{s,c_{opt}}$ can be chosen such that its coherent sum with the fixed terms $\bar{\bar{G}}_0 + \langle \bar{\bar{G}}_s \rangle_c$ yields the overlap between the port's mode and the chaotic cavity's modes that is necessary to adjust the coupling rate to the required value. The required value is the chaotic cavity's intrinsic decay rate which may in principle also depend on how the cavity's boundary conditions are configured, especially if strong localized loss mechanisms are present which allow the wave's exposure to these losses to be tailored via the boundary conditions. In our experiment, however, global attenuation effects dominate. The effect of the coherent interplay of the Green's function components can be observed in **Supplementary Figure 7**: the free-space reflection (black) corresponds to $\bar{\bar{G}} = \bar{\bar{G}}_0$, the reflection averaged over configurations (purple) corresponds to $\bar{\bar{G}} = \bar{\bar{G}}_0 + \langle \bar{\bar{G}}_s \rangle_c$, the reflection with a random metasurface configuration (red) to $\bar{\bar{G}} = \bar{\bar{G}}_0 + \langle \bar{\bar{G}}_s \rangle_c + \bar{\bar{G}}_{s,c_{rand}}$ and the lowest achievable reflection (blue) corresponds to $\bar{\bar{G}} = \bar{\bar{G}}_0 + \langle \bar{\bar{G}}_s \rangle_c + \bar{\bar{G}}_{s,c_{opt}}$, where the optimal configuration $c_{opt}$ for $f_0 = 20.52$ GHz is chosen. At some frequencies the scattering contribution reduces the reflected power relative to free space while at other frequencies it has the opposite effect, depending on how $\bar{\bar{G}}_0$ and $\bar{\bar{G}}_s$ add coherently. Therefore, by controlling $\bar{\bar{G}}_{s,c}$ we could also maximize rather than minimize the reflection amplitude.

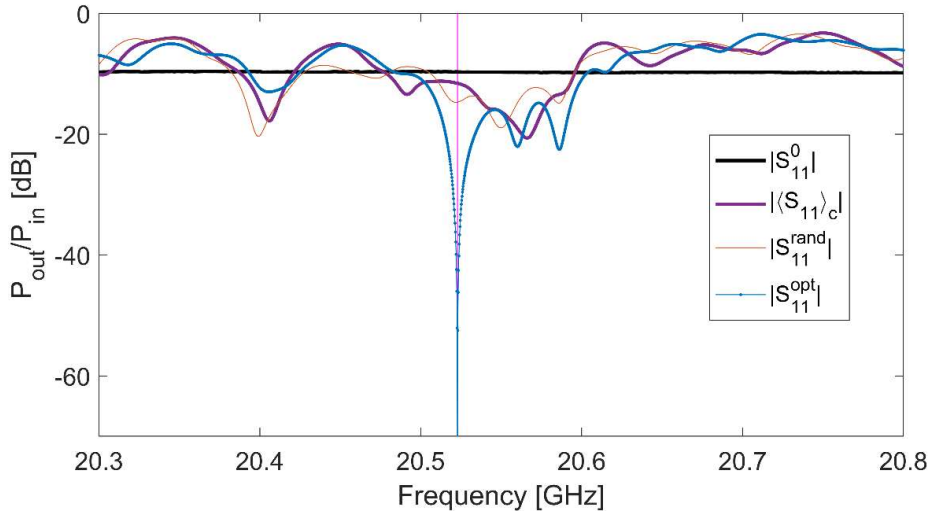

**Supplementary Figure 7. Reflected power spectra for various constellations.** Reflected power spectra averaged over metasurface configurations (purple), for a random metasurface configuration (red) and for the optimal metasurface configuration to achieve PA at $f_0 = 20.52$ GHz (blue) are shown. We also plot the reflected power spectrum measured with the port in an anechoic environment (black).

# Supplementary Note 7. Details on Receiver-Powered Secure Wireless Communication Protocol

In this section, we provide additional details on the proposal and demonstration of receiver-powered secure wireless communication presented in the main text. First, Alice and Bob agree on a set of frequencies for which Alice is capable of imposing a PA condition on Bob's port. In **Figure 3a**, eight such frequencies are chosen and indicated with arrows. This defines a first set of eight metasurface configurations that Alice can use to impose a PA condition on Bob's port. Next, Alice picks another set of metasurface configurations, for which PA does not occur at any of the selected frequencies on Bob's port.

In order to transfer a binary piece of information $b$ to Bob, Alice begins by picking a random integer $n$ such that $1 \leq n \leq 8$, 8 being the number of selected frequencies in our example. Then, if the bit to be sent is $b = 1$, Alice configures the metasurface such that she imposes PA on Bob's port for the $n$th selected frequency. If $b = 0$, Alice picks the $n$th random configuration that does not yield PA at any of the selected frequencies on Bob's port. In order to decode the piece of information sent by Alice, Bob measures the reflected power at his port at all of the eight selected frequencies. If for one of those frequencies the reflected power is below a certain threshold value (e.g. -60 dB), Bob decodes $b = 1$. Otherwise, Bob concludes that $b = 0$. This scheme is summarized in **Supplementary Figure 8**.

The security of this scheme relies on (i) the extreme sensitivity of the PA condition, and (ii) the ease of achieving PA at multiple frequencies with our technique. The extreme sensitivity ensures that PA can only be observed on Bob's port but not on Eve's port. Moreover, if an eavesdropper or manipulator appears during communication, the system is detuned and the PA condition is destroyed due to its extreme sensitivity, such that Alice and Bob notice any intrusion. The ability to easily achieve PA at multiple frequencies ensures that Eve cannot identify the transmitted message up to an uncertainty about which configuration is a "1" and which a "0". Indeed, although Eve cannot observe PA on her port, with a high dynamic range she can certainly detect that the system alternates between two configurations according to the transmitted bit series. If, however, Alice uses a different PA/nonPA configuration for each bit, Eve cannot make such an observation, she only receives random numbers. This is evidenced with *in situ* measurements in **Figure 3c**.

Finally, we highlight the unique "receiver-powered" character of our scheme. Indeed, Alice does not emit any waves, nor does she rely on the presence of ambient stray waves. She only configures the system's geometry with the programmable metasurface which requires a negligible amount of energy[29]. Bob, despite being the receiver, is the one who actively emits waves in order to measure his port's reflection spectrum. Our scheme therefore offers the unique possibility for Alice to transfer information (i) at high security, and (ii) with negligible power consumption.

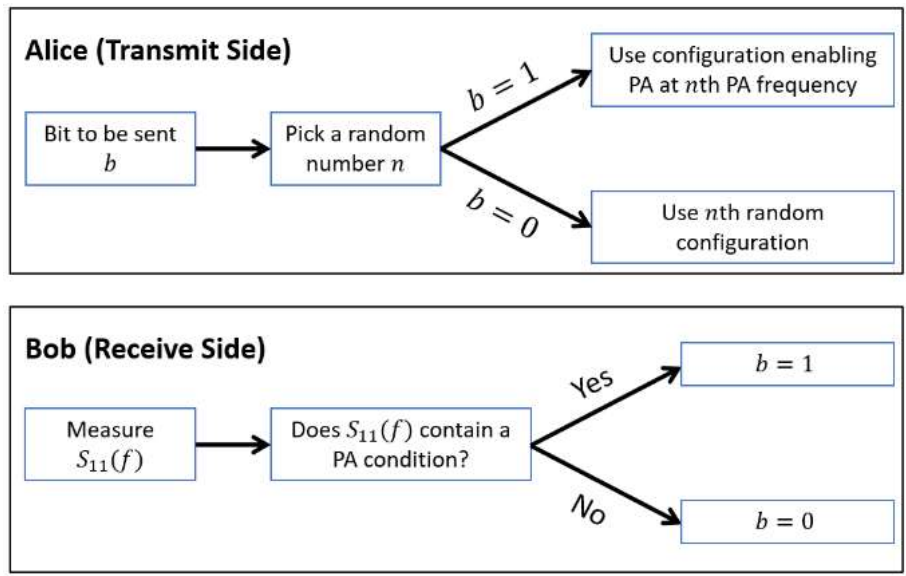

**Supplementary Figure 8.** **Block diagram for receiver-powered secure wireless communication.** Alice and Bob agree on $n$ frequencies at which Alice may impose PA on Bob's port (in our example, $n = 8$). For every bit that Alice sends, she randomly picks one out of these $n$ selected frequencies. This procedure ensures that there is more than one metasurface configuration corresponding to each possible bit value, such that an eavesdropper Eve cannot identify if the value of two consecutive bits is identical or not.